\documentclass{ws-ijmpd}
\usepackage{amsmath}
\usepackage{amssymb}

\allowdisplaybreaks

\begin{document}

\title{Modified Jordan-Brans-Dicke theory with scalar current and the Eddington-Robertson $\gamma$-parameter}

\author{J. W. Moffat$^{\star,\dagger}$ and V. T. Toth$^\star$\\~\\
{\rm
\footnotesize
$^\star$Perimeter Institute for Theoretical Physics, Waterloo, Ontario N2L 2Y5, Canada\\
$^\dagger$Department of Physics, University of Waterloo, Waterloo, Ontario N2L 3G1, Canada}}

\maketitle

\begin{abstract}
The Jordan-Brans-Dicke theory of gravitation, which promotes the gravitational constant to a dynamical scalar field, predicts a value for the Eddington-Robertson post-Newtonian parameter $\gamma$ that is significantly different from the general relativistic value of unity. This contradicts precision solar system measurements that tightly constrain $\gamma$ around 1. We consider a modification of the theory, in which the scalar field is sourced explicitly by matter. We find that this leads to a modified expression for the $\gamma$-parameter. In particular, a specific choice of the scalar current yields $\gamma=1$, just as in general relativity, while the weak equivalence principle is also satisfied. This result has important implications for theories that mimic Jordan-Brans-Dicke theory in the post-Newtonian limit in the solar system, including our scalar-tensor-vector modified gravity theory (MOG).
\end{abstract}

%\pacs{04.20.Cv,04.50.Kd,04.80.Cc,98.80.-k}

%\maketitle

~\par

Jordan-Brans-Dicke\cite{Jordan1955,BD1962} theory is a theory of gravitation in which the gravitational constant $G$ is replaced with the inverse of a dynamical scalar field $\phi$. It can be demonstrated by straightforward derivation that this scalar field is effectively sourced by the curvature of space-time\cite{Weinberg1972}. There is, however, no scalar current: in the Lagrangian formulation, the variation of matter fields with respect to the scalar field is assumed to be zero.

Many theories predict the same phenomenology as Jordan-Brans-Dicke theory. These include $f(R)$ theories of gravity\cite{Sotiriou2006}, quintessence/dark energy models\cite{Bertolami2000,Torres2002}, inflaton fields\cite{LaDaile1989}, and superstring theory\cite{GSW1987}, as well as our scalar-tensor-vector (STVG) modified gravity theory (MOG\cite{Moffat2006a,Moffat2007e}). However, Jordan-Brans-Dicke theory runs into severe observational constraints within the solar system. Notably, the theory predicts that the value of the post-Newtonian $\gamma$-parameter, first introduced by Eddington\cite{Eddington} and Robertson\cite{Robertson1938} and also Schiff\cite{Schiff1960a}, and effectively measuring the amount of spatial curvature produced by unit rest mass, will deviate from the standard general relativistic value of 1\footnote{The other Eddington-Robertson parameter, $\beta$, is identically 1 in Jordan-Brans-Dicke theory, just as in general relativity.}. Instead, its value will be\cite{Weinberg1972} $\gamma=(\omega+1)/(\omega+2)$, where $\omega$ is the dimensionless coupling constant of the dynamical field. Constraints established by precision measurements of the Cassini spacecraft\cite{BT2003} require the uncomfortably large value of $|\omega|>4\times 10^4$.

On the other hand, there is no {\em a priori} reason to exclude the possibility of a scalar current. Indeed, one might argue that it is the assumption that matter does not couple directly to the scalar field that is an artificially imposed restriction. In microphysics, scalar fields such as the Higgs field are routinely assumed to couple to matter\cite{Burgess2007}. The coupling of scalar fields to matter has also been considered, for instance, in attempts to formulate MOND-like gravity as a field theory\cite{Bruneton2007}, or in the context of extended inflation\cite{Holman1990}.

A phenomenological matter Lagrangian could be constructed such that it depends explicitly on $G=\phi^{-1}$. The variation of such a Lagrangian with respect to $\phi$ would be non-zero, introducing a scalar current into the field equations. To demonstrate this, we write the scalar theory Lagrangian as follows:
\begin{equation}
{\cal L}=\frac{1}{16\pi}\left[(R - 2\Lambda)\phi + f(\phi,g^{\mu\nu}\partial_\mu\phi\partial_\nu\phi)\right]\sqrt{-g} + {\cal L}_\mathrm{O.F.},\label{eq:L}
\end{equation}
where $R$ is the Ricci-scalar constructed from the metric $g_{\mu\nu}$, $g$ is the metric determinant, $\Lambda$ is the cosmological constant, $\phi$ is a scalar field, $f$ is an arbitrary function, and O.F. stands for terms that represent other fields. We assume that these other fields depend only on $\phi$, not on its derivatives. We set $c=1$, use the $(+,-,-,-)$ metric signature, and define the Ricci tensor as $R_{\mu\nu}=\partial_\alpha\Gamma^\alpha_{\mu\nu}-\partial_\nu\Gamma^\alpha_{\mu\alpha}+\Gamma^\alpha_{\mu\nu}\Gamma^\beta_{\alpha\beta}-\Gamma^\alpha_{\mu\beta}\Gamma^\beta_{\alpha\nu}$, where the $\Gamma$ are the usual Christoffel-symbols.

The field equations of the theory are the Euler-Lagrange equations corresponding to (\ref{eq:L}):
\begin{align}
\frac{\partial{\cal L}}{\partial g^{\mu\nu}}-\partial_\kappa\frac{\partial{\cal L}}{\partial g^{\mu\nu}_{,\kappa}}+\partial_{\kappa}\partial_{\lambda}\frac{\partial{\cal L}}{\partial g^{\mu\nu}_{,\kappa\lambda}}=&0,\\
\frac{\partial{\cal L}}{\partial\phi}-\nabla_\mu\frac{\partial{\cal L}}{\partial(\partial_\mu\phi)}=&0,
\end{align}
where $\nabla_\mu$ is the covariant derivative with respect to $x^\mu$. These equations can be recast in the form,
\begin{align}
R_{\mu\nu}-\frac{1}{2}g_{\mu\nu}R+g_{\mu\nu}\Lambda+\frac{1}{\sqrt{-g}}\frac{1}{\phi}\frac{\partial f\sqrt{-g}}{\partial g^{\mu\nu}}-\frac{\partial R}{\partial g^{\mu\nu}_{,\kappa}}\frac{\partial_\kappa\phi}{\phi}&\nonumber\\
+\frac{2}{\sqrt{-g}}\partial_\lambda\left(\sqrt{-g}\frac{\partial R}{g^{\mu\nu}_{,\kappa\lambda}}\right)\frac{\partial_\kappa\phi}{\phi}+\frac{\partial R}{\partial g^{\mu\nu}_{,\kappa\lambda}}\frac{\partial_\kappa\partial_\lambda\phi}{\phi}&=\frac{8\pi}{\phi}T_{\mu\nu},\label{eq:EFE}\\
R-2\Lambda+\frac{\partial f}{\partial\phi}-\nabla_\kappa\frac{\partial f}{\partial(\partial_\kappa\phi)}&=16\pi J\label{eq:EQG},
\end{align}
where $T_{\mu\nu}=-(2/\sqrt{-g})\partial{\cal L}_\mathrm{O.F.}/\partial g^{\mu\nu}$ and $J=-(1/\sqrt{-g})\partial{\cal L}_\mathrm{O.F.}/\partial\phi$. The existence of a non-zero variation of matter fields with respect to $\phi$ represents a significant generalization of the archetypal scalar field theory of Jordan, Brans and Dicke.

Equation (\ref{eq:EFE}) can be rewritten using covariant derivatives, yielding
\begin{equation}
R_{\mu\nu}-\frac{1}{2}g_{\mu\nu}R+g_{\mu\nu}\Lambda+\frac{1}{\sqrt{-g}}\frac{1}{\phi}\frac{\partial f\sqrt{-g}}{\partial g^{\mu\nu}}-\frac{\partial R}{\partial g^{\mu\nu}_{\kappa\lambda}}\frac{\nabla_\kappa\nabla_\lambda\phi}{\phi}=\frac{8\pi}{\phi}T_{\mu\nu},
\end{equation}
Spelled out, the field equations now take the following form:
\begin{align}
\hskip -12pt
R_{\mu\nu}-\frac{1}{2}g_{\mu\nu}R+g_{\mu\nu}\Lambda+w\frac{\partial_\mu\phi\partial_\nu\phi}{\phi}-\frac{1}{2}g_{\mu\nu}\frac{f}{\phi}
+(g^{\kappa\lambda}g_{\mu\nu}-\delta^\kappa_\mu\delta^\lambda_\nu)\frac{\nabla_\kappa\nabla_\lambda\phi}{\phi}
=&\frac{8\pi}{\phi}T_{\mu\nu},\label{eq:EFE1}\\
R-2\Lambda+v-2\partial_\mu w\partial^\mu\phi-2w\nabla_\mu\nabla^\mu\phi=&16\pi J,\label{eq:EQG1}
\end{align}
where $v=\partial f/\partial\phi$ and $w=\partial f/\partial(g^{\mu\nu}\partial_\mu\phi\partial_\nu\phi)$. Taking the trace of (\ref{eq:EFE1}), we obtain
\begin{equation}
-R+4\Lambda+w\frac{\partial_\mu\phi\partial^\mu\phi}{\phi}-2\frac{f}{\phi}+3\frac{\nabla_\mu\nabla^\mu\phi}{\phi}=\frac{8\pi}{\phi}T.
\end{equation}
This allows us to rewrite (\ref{eq:EFE1}) and (\ref{eq:EQG1}) as
\begin{align}
R_{\mu\nu}=&\frac{8\pi}{\phi}\left\{T_{\mu\nu}+\frac{1}{3-2w\phi}\left[\phi J-(1-w\phi)\left(T-\frac{1}{4\pi}\phi\Lambda\right)\right]g_{\mu\nu}\right\}\nonumber\\
&{}+\frac{1-w\phi}{3-2w\phi}\left(w\partial_\mu\phi\partial^\mu\phi-\frac{1}{2}f\right)g_{\mu\nu}-\frac{1}{3-2w\phi}\left(\frac{1}{2}v-\partial_\mu w\partial^\mu\phi\right)g_{\mu\nu}\nonumber\\
&{}+\frac{\nabla_\mu\nabla_\nu\phi}{\phi}-\frac{w\partial_\mu\phi\partial_\nu\phi}{\phi},\\
\nabla_\mu\nabla^\mu\phi=&\frac{2}{3-2w\phi}\left(16\pi T-4\phi\Lambda+32\pi\phi J+4f-2v\phi-2w\partial_\mu\phi\partial^\mu\phi+4\partial_\mu w\partial^\mu\phi\right).
\end{align}

For Jordan-Brans-Dicke theory, $f(\phi, g^{\mu\nu}\partial_\mu\phi\partial_\nu\phi)=-\omega\partial_\mu\phi\partial^\mu\phi/\phi$, hence $v=-f/\phi$ and $w=-\omega/\phi$. Therefore, the equations read
\begin{align}
R_{\mu\nu}=&\frac{8\pi}{\phi}\left\{T_{\mu\nu}+\frac{1}{2\omega+3}\left[\phi J-(\omega+1)\left(T-\frac{1}{4\pi}\phi\Lambda\right)\right]g_{\mu\nu}\right\}+\omega\frac{\partial_\mu\phi\partial_\nu\phi}{\phi^2}\nonumber\\
&{}+\frac{\nabla_\mu\nabla_\nu\phi}{\phi},\label{eq:EFE2}\\
\nabla_\mu\nabla^\mu\phi=&\frac{8\pi}{2\omega+3}\left(T+2\phi J-\frac{1}{4\pi}\phi\Lambda\right),\label{eq:EQG2}
\end{align}
which, apart from the presence of $J$, are the equations of Jordan-Brans-Dicke theory in the standard form. To the first post-Newtonian order, terms quadratic in derivatives vanish; the second derivative in (\ref{eq:EFE2}) can, in turn, be eliminated by a suitable gauge choice (for a thorough derivation, see Appendix A of Ref.~\refcite{Deng2009}). In the post-Newtonian metric\cite{Will1993}, $T\simeq T_{00}$ and the $\gamma$-parameter can be read off as the ratio of the $ii$ and $00$ components of (\ref{eq:EFE2}). In the absence of a cosmological term, $\Lambda=0$, we get
\begin{equation}
\gamma=\frac{(\omega+1)T-\phi J}{(\omega+2)T+\phi J}
\end{equation}

If the scalar current vanishes ($J=0$), we get back the usual post-Newtonian result for Jordan-Brans-Dicke theory:
\begin{equation}
\gamma=\frac{\omega+1}{\omega+2}.\label{eq:BDgamma}
\end{equation}
This result is frequently cited as a reason for rejecting Jordan-Brans-Dicke theory within the solar system, as precision measurements by the Cassini spacecraft yielding $\gamma-1=(2.1\pm 2.3)\times 10^{-5}$, for instance, are consistent with the theory only if\cite{BT2003} $|\omega|\gtrsim 4\times 10^{4}$.

However, if a scalar current is present, the situation changes. Specifically, we can choose a scalar current in the form
\begin{equation}
\phi J=-\frac{1}{2}T,\label{eq:WEPcond}
\end{equation}
which is equivalent to
\begin{equation}
-\phi\frac{1}{\sqrt{-g}}\frac{\partial{\cal L}_\mathrm{O.F.}}{\partial\phi}=\frac{1}{\sqrt{-g}}\frac{\partial{\cal L}_\mathrm{O.F.}}{\partial g^{\mu\nu}}g^{\mu\nu}.
\end{equation}
This choice can be made, in part, because $J$ is not a conserved quantity, just as $T$ is not conserved. In this case, equations (\ref{eq:EFE2}) and (\ref{eq:EQG2}) read
\begin{align}
R_{\mu\nu}=&\frac{8\pi}{\phi}\left(T_{\mu\nu}-\frac{1}{2}Tg_{\mu\nu}+\frac{1}{4\pi}\frac{\omega+1}{2\omega+3}\phi\Lambda g_{\mu\nu}\right)+\omega\frac{\partial_\mu\phi\partial_\nu\phi}{\phi^2}+\frac{\nabla_\mu\nabla_\nu\phi}{\phi},\label{eq:EFE3}\\
\nabla_\mu\nabla^\mu\phi=&-\frac{2\phi\Lambda}{2\omega+3}.\label{eq:EQG3}
\end{align}
Considering the trace of the bracketed term in Eq.~(\ref{eq:EFE3}), if
\begin{equation}
|\Lambda|\ll \pi\left|\frac{2\omega+3}{\omega+1}\phi^{-1}T\right|,
\end{equation}
the general relativistic result that is also consistent with solar system data,
\begin{equation}
\gamma\simeq 1,\label{eq:GRgamma}
\end{equation}
is easily satisfied.

The condition (\ref{eq:WEPcond}) is satisfied trivially by minimally coupled scalar-tensor theories, in which case the coupling between the scalar field and geometry on the one hand, and matter on the other can be simultaneously transformed away with a conformal transformation, leaving only kinetic scalar terms in the Lagrangian. This confirms the known fact that these theories are consistent with the limits established by solar system and E\"otv\"os-type experiments.

For a general (scalar, spinor, vector, tensor) field theory, similar considerations apply. The condition (\ref{eq:WEPcond}) can be satisfied by a trivial coupling, in which case a conformal transformation transforms the theory into one in which the Jordan-Brans-Dicke type scalar field couples only minimally to other forms of matter. However, the solution space is not restricted to the minimally coupled theory. One may postulate a trial Lagrangian in the form ${\cal L}_\mathrm{O.F.}=f(\phi){\cal L}_\mathrm{matter}+h(\phi,\partial_\alpha\phi\partial^\alpha\phi)$, where ${\cal L}_\mathrm{matter}$ represents matter fields before the application of the Jordan-Brans-Dicke type field. One can then solve for $f(\phi)$ and $h(\phi,\partial_\alpha\phi\partial^\alpha\phi)$ using (\ref{eq:WEPcond}) and possible additional conditions, such as demanding that $T_\mathrm{O.F.}$, obtained by varying ${\cal L}_\mathrm{O.F.}$ with respect to $g^{\mu\nu}$, be a constant multiple of $T_\mathrm{matter}$, obtained from the variation of ${\cal L}_\mathrm{matter}$.

The result (\ref{eq:BDgamma}) has been used as an argument against theories that, within the solar system, yield the same solution as Jordan-Brans-Dicke theory to the first post-Newtonian order. We mention in particular our scalar-tensor-vector (STVG) modified gravity theory (MOG\cite{Moffat2006a,Moffat2007e}), which, according to an extensive analysis by Deng, et al\cite{Deng2009}, shows the same behavior in the solar system as Jordan-Brans-Dicke theory. This problem is avoided by a suitable choice of $J$ yielding (\ref{eq:GRgamma}), as demonstrated above.

Nonetheless, we note that in the case of $J\ne 0$, the theory is no longer a metric theory: massive particles carry a scalar charge and no longer move along geodesics. To determine the equations of motion for a test particle, we use a test particle Lagrangian in the form
\begin{equation}
L_\mathrm{TP}=-m\sqrt{g_{\mu\nu}u^\mu u^\nu}-q\phi,\label{eq:TPL}
\end{equation}
where $q$ is the scalar charge associated with a particle of mass $m$, moving with four-velocity $u^\mu=dx^\mu/d\tau$ and $\tau$ is the proper time along the particle's world line. Integration of (\ref{eq:WEPcond}) over a three-volume encompassing a test particle gives
\begin{equation}
q=-\frac{1}{2}\phi^{-1}m,\label{eq:q}
\end{equation}
and $\frac{1}{2}\phi^{-1}m\simeq\frac{1}{2}G_Nm$ at the present epoch ($G_N$ is Newton's constant of gravitation.) The equation of motion obtained by varying (\ref{eq:TPL}) contains an extra term when compared to the standard geodesic equation of motion:
\begin{equation}
m\left(\frac{d^2x^\kappa}{d\tau^2}+\Gamma_{\mu\nu}^\kappa u^\mu u^\nu\right)-qg^{\kappa\lambda}\frac{\partial\phi}{\partial x^\lambda}=0.
\end{equation}
Given (\ref{eq:q}), we obtain
\begin{equation}
m\left(\frac{d^2x^\kappa}{d\tau^2}+\Gamma_{\mu\nu}^\kappa u^\mu u^\nu\right)+mg^{\kappa\lambda}\frac{1}{2\phi}\frac{\partial\phi}{\partial x^\lambda}=0.\label{eq:motion}
\end{equation}
We observe that $m$ cancels out in the equation of motion. Hence, the equivalence of passive gravitational and inertial mass\cite{Will2006} is automatically satisfied. The acceleration of a test particle is given by

\begin{equation}
\ddot x^\kappa = -\Gamma_{\mu\nu}^\kappa u^\mu u^\nu - \frac{1}{2}\partial^\kappa\ln\phi,
\end{equation}
where $\phi$ is determined by the homogeneous source-free equation (\ref{eq:EQG3}). The presence of the fifth-force term in this equation may lead to deviations from the predictions of Einstein gravity. Furthermore, we note that only massive particles are subject to the fifth force; massless photons continue to travel on light-like geodesics. These predictions can be contrasted with experimental results in the solar system and be used to constrain the solution space for $\phi$.

Finally, we note that equation (\ref{eq:EQG3}) can be rewritten in the familiar form
\begin{equation}
(\Box+\mu^2)\phi=0,
\end{equation}
with $\Box=\nabla_\nu\nabla^\nu$ and $\mu$ given by
\begin{equation}
\mu^2=\frac{2\Lambda}{2\omega+3}.
\end{equation}
This last term can be interpreted as the mass $\mu$ of the scalar field $\phi$. Using $\Lambda\simeq 1.2\times 10^{-52}$~m$^{-2}$, we obtain the mass of an ultralight scalar field, $\mu\simeq 3.9\sqrt{2/(2\omega+3)}\times 10^{-69}$~kg.

Our research does not directly imply that scalar tensor theories may be able to account for cosmological observations such as weak lensing\cite{Schimd2005}. However, it vindicates theories that are able to deal with such observations, yet have been criticized or rejected because in the solar system, they are unable to account for the stringent observational limits on the $\omega$ parameter.

\section*{Acknowledgments}

JWM thanks the John Templeton Foundation for generous financial support. The research was partially supported by National Research Council of Canada. Research at the Perimeter Institute for Theoretical Physics is supported by the Government of Canada through NSERC and by the Province of Ontario through the Ministry of Research and Innovation (MRI).

\bibliography{refs}
\bibliographystyle{unsrt}

\end{document}